\documentclass[conference]{IEEEtran}
\IEEEoverridecommandlockouts
\usepackage{cite}
\usepackage{amsmath,amssymb,amsfonts}
\usepackage{algorithmic}
\usepackage{graphicx}
\usepackage{textcomp}
\usepackage{xcolor}
\usepackage{multirow}
\usepackage{multicol}
\usepackage{subfigure}
\def\BibTeX{{\rm B\kern-.05em{\sc i\kern-.025em b}\kern-.08em
    T\kern-.1667em\lower.7ex\hbox{E}\kern-.125emX}}
\begin{document}

\title{AsCL: An Asymmetry-sensitive Contrastive Learning Method for Image-Text Retrieval with Cross-Modal Fusion
}

\author{\IEEEauthorblockN{1\textsuperscript{st} Ziyu Gong}
\IEEEauthorblockA{
\textit{State Key Laboratory for} \\
\textit{Novel Software Technology}\\
\textit{Nanjing University}\\
Nanjing, China \\
ziyugong@smail.nju.edu.cn}
\and
\IEEEauthorblockN{2\textsuperscript{nd} Chengcheng Mai}
\IEEEauthorblockA{
\textit{School of Computer Science and Electronic Information/} \\
\textit{School of Artifical Intelligence}\\
\textit{Nanjing Normal University}\\
Nanjing, China \\
maicc@njnu.edu.cn}
\and
\IEEEauthorblockN{3\textsuperscript{rd} Yihua Huang*}
\IEEEauthorblockA{
\textit{State Key Laboratory for} \\
\textit{Novel Software Technology}\\
\textit{Nanjing University}\\
Nanjing, China \\
yhuang@nju.edu.cn}
}

\maketitle

\begin{abstract}
The image-text retrieval task aims to retrieve relevant information from a given image or text. The main challenge is to unify multimodal representation and distinguish fine-grained differences across modalities, thereby finding similar contents and filtering irrelevant contents. However, existing methods mainly focus on unified semantic representation and concept alignment for multi-modalities, while the fine-grained differences across modalities have rarely been studied before, making it difficult to solve the information asymmetry problem. In this paper, we propose a novel asymmetry-sensitive contrastive learning method. By generating corresponding positive and negative samples for different asymmetry types, our method can simultaneously ensure fine-grained semantic differentiation and unified semantic representation between multi-modalities. Additionally, a hierarchical cross-modal fusion method is proposed, which integrates global and local-level features through a multimodal attention mechanism to achieve concept alignment. Extensive experiments performed on MSCOCO and Flickr30K, demonstrate the effectiveness and superiority of our proposed method. 
\end{abstract}


\section{Introduction}
Image-text retrieval aims to search for relevant text based on visual queries and vice versa. Early efforts contribute significantly to the learning of unified multimodal representation and visual-textual concept alignment, but it still faces the following challenge: the existing work rarely considers the problem of cross-modal information asymmetry, which neglects the fine-grained differences between modalities.

The information asymmetry problem in the multimodal retrieval task refers to the unequal information capacity between different modalities, \textit{i.e.}, when describing the same scene, one modality may contain more or less information than the other modality. Image modality objectively describes the scene based on pixels, while text modality describes the scene based on characters or words. It should be noted that different modalities have different emphases on the description of a certain scene. For a certain image, the paired text may be partially similar to the image and partially different from the image. The information asymmetry problem aggravates the difficulty of the image-text retrieval task.
\begin{figure}
\centering
\includegraphics[width=1\linewidth]{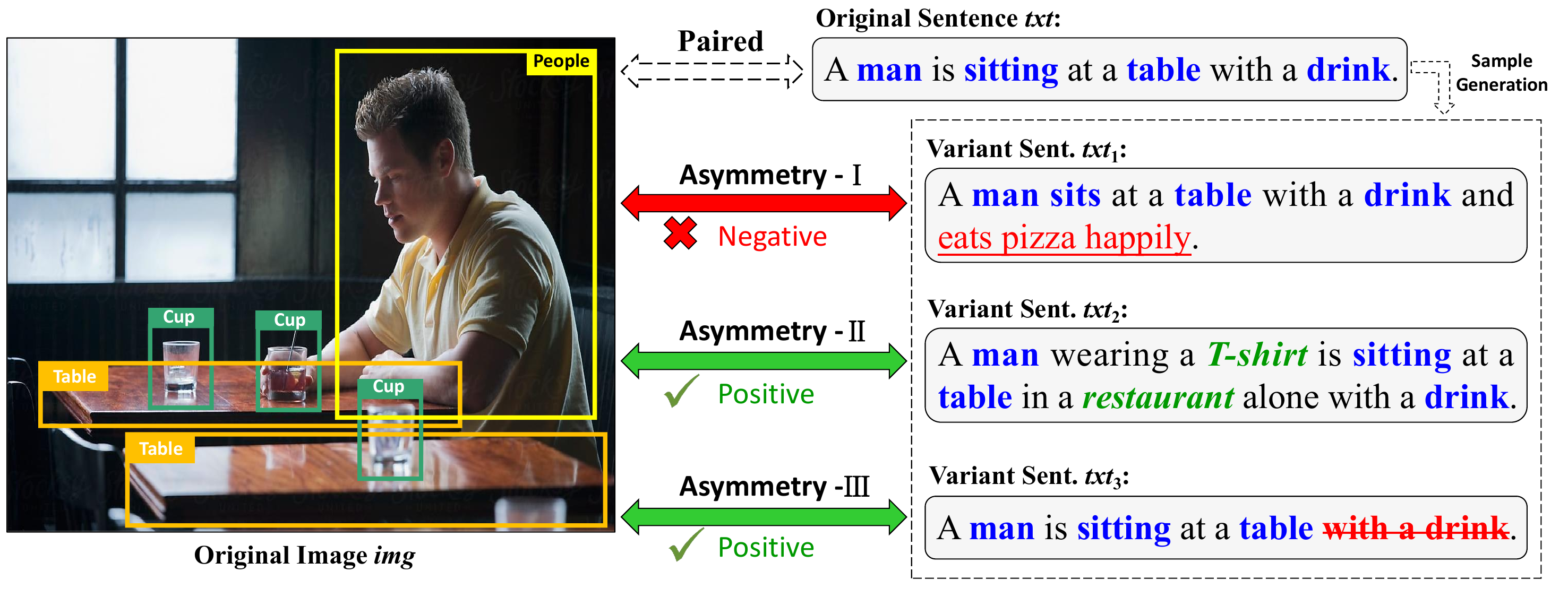}
\caption{Fine-grained information asymmetry types.} 
\label{Asymmetry}
\end{figure}
As shown in Figure \ref{Asymmetry}, for a given paired image-text pair, (\textit{img}, \textit{txt}), we subdivided information asymmetry into three types. (1) Asymmetry-\uppercase\expandafter{\romannumeral1}: The text contains redundant information that does not exist in the image. For variant \textit{$txt_1$}, although \textit{$txt_1$} contains most of the objects corresponding to image \textit{img}, such as ``people'', ``table'', and ``cup'', \textit{$txt_1$} mistakenly contains ``pizza'' that is not mentioned in image. Therefore, \textit{$txt_1$} should be considered a negative of \textit{img}. However, existing models would mistakenly retrieve \textit{$txt_1$} as a positive because it contains most of the information in the image. (2) Asymmetry-\uppercase\expandafter{\romannumeral2}: Compared to the original paired text, the variant text contains more relevant information belonging to the corresponding image. For variant \textit{$txt_2$}, it is a more informative illustration of \textit{img}, which contains more details about clothing and location without irrelevant content, and should be regarded as a positive of \textit{img}. (3) Asymmetry-\uppercase\expandafter{\romannumeral3}: Compared to the original paired text, the variant text discards partial information contained in the given image, but still conforms to the description of the image. For variant \textit{$txt_3$}, although the word “drink” has been deleted, it should still be considered a positive of \textit{img} as it fits the description of \textit{img}. However, existing models may mistakenly identify it as a negative or lower its retrieval ranking because the proportion of overlapping concepts between \textit{$txt_3$} and \textit{img} has decreased. 

To address the above issues, a novel asymmetry-sensitive contrastive learning method was presented. For each fine-grained information asymmetry type, we generated corresponding positive or negative textual samples with both partially similar parts and subtly different parts, which can be leveraged in the optimization of contrastive learning and acquire more discriminative multimodal semantic representations. In particular, for Asymmetry-\uppercase\expandafter{\romannumeral1}, we generated negative samples by adding noise information into the embedding layer to fully enhance the diversity of the generated negatives. For Asymmetry-\uppercase\expandafter{\romannumeral2} and Asymmetry-\uppercase\expandafter{\romannumeral3}, we generated positive samples by concatenating and truncating keywords on the original input statements, respectively, in order to enrich the diversity of semantic descriptions for the same scenario. 

Our method enhanced the sensitivity of the contrastive learning algorithm to the fine-grained information asymmetry across modalities by increasing the diversity of positive and negative samples, and provided semantic representation support for the subsequent multimodal retrieval task based on semantic similarity. In addition, in order to capture more sophisticated correlations across modalities, a hierarchical cross-modal fusion has also been proposed, for achieving multimodal concept alignment at both image-text and region-word levels, through a cross-modal attention mechanism. 

The major contributions can be concluded as follows: (1) An asymmetry-sensitive contrastive learning method was proposed to solve the fine-grained information asymmetry problem between images and texts, where corresponding positives and negatives for each asymmetry type are generated to achieve unified semantic representation for better cross-modality retrieval based on semantic similarity. (2) We also presented an image-text feature fusion and semantic alignment method based on a cross-modal attention mechanism for high-quality modality interaction, from both global and local perspectives. (3) Extensive experiments verified that our method outperformed the state-of-the-art baselines.

\section{Our AsCL Methodology}
\begin{figure*}[ht]
\centering
\includegraphics[width=0.8\linewidth]{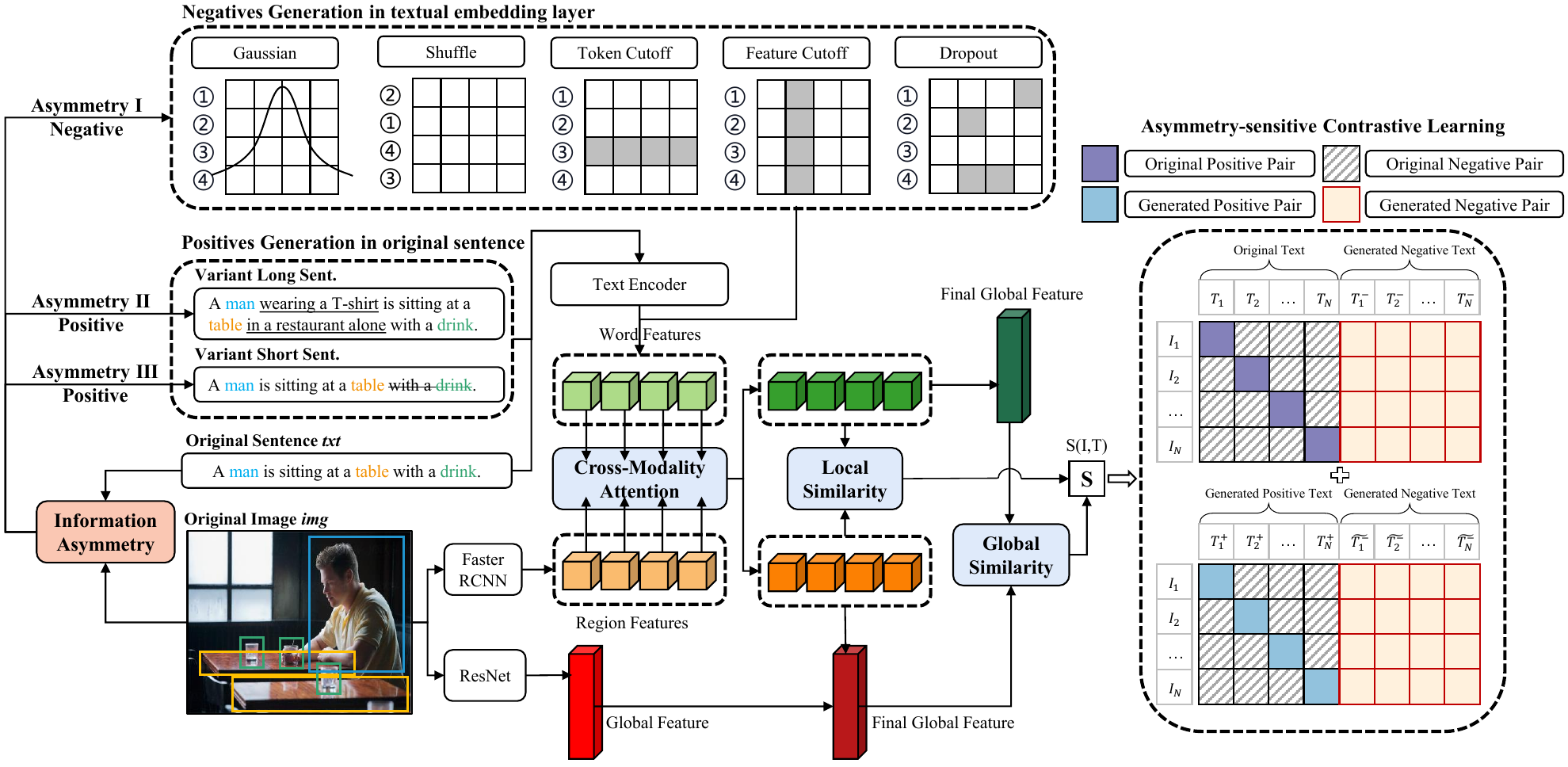}
\caption{The overview of our proposed AsCL Method.} \label{fig:Architecture}
\end{figure*}

\subsection{Multi-modal Feature Representation}
As shown in Figure \ref{fig:Architecture}, for each input image, we extracted $K$ region features with the pretrained Faster-RCNN model and then projected them into a $D$-dimensional space, denoted as $ \boldsymbol{V} = [\boldsymbol{v}_{1},\boldsymbol{v}_{2},...,\boldsymbol{v}_{K} ] \in \mathbb{R}^{K*D}$. Meanwhile, we acquired the global feature with the pretrained ResNet152 model and then transformed it into a $D$-dimensional vector, \textit{i.e.}, $\boldsymbol{G} \in \mathbb{R}^{D}$.

For each input text, we utilized BERT to obtain word embeddings, $\boldsymbol{W} = [ \boldsymbol{w}_{1}, \boldsymbol{w}_{2},...,\boldsymbol{w}_{L} ] \in \mathbb{R}^{L*D}$, where $L$ is the number of words and $D$ is the dimension of word features.

\subsection{Information Asymmetry and Sample Generation}
Information asymmetry quantifies the differences between the sets of image and text information. According to fine-grained information asymmetries, we generated different samples. \textbf{Asymmetry-\uppercase\expandafter{\romannumeral1}}: The text contains more redundant information that does not belong to the corresponding image. We used four methods to generate diverse negative samples by adding noise information to the textual embedding layers: (\romannumeral1) Gaussian Noise: Gaussian noise is random variation that follows a Gaussian (normal) distribution. We directly attached it to $\boldsymbol{W}$. (\romannumeral2) Token Shuffling: The order of tokens in $\boldsymbol{W}$ was randomly altered. (\romannumeral3) Token Cutoff \& Feature Cutoff: We assigned zero to a row (token) or a column (feature) of $\boldsymbol{W}$. (\romannumeral4) Dropout: We randomly discarded elements in $\boldsymbol{W}$ based on a specific probability. \textbf{Asymmetry-\uppercase\expandafter{\romannumeral2}}: The text contains more relevant information that belongs to the corresponding image. For each image in the original dataset, there are five corresponding captions sharing similar concepts. We generated a long positive sentence by randomly selecting two related sentences and concatenating them into the long sentence. \textbf{Asymmetry-\uppercase\expandafter{\romannumeral3}}: The text discards partial relevant information that belongs to the corresponding image. We truncated the original text to generate a short but positive sentence, which increases the diversity of positives.

\subsection{Cross-modal Fusion}
\textbf{Local Region-Word Fusion.} Due to the bidirectional requirement of image-text retrieval, we defined $\boldsymbol{X} = [\boldsymbol{x}_1,\boldsymbol{x}_2,...,\boldsymbol{x}_M] \in \mathbb{R}^{M*D}$ and $\boldsymbol{Y} = [\boldsymbol{y}_1,\boldsymbol{y}_2,...,\boldsymbol{y}_N] \in \mathbb{R}^{N*D}$ to denote the given query (image or text) and the retrieved results (text or image), respectively, where $\boldsymbol{x}_{i} \in \mathbb{R}^{D}$ and $\boldsymbol{y}_{i} \in \mathbb{R}^{D}$ are represented as region-level or word-level features. In this work, we implemented two single symmetrical versions of formula. Hence, we set $\boldsymbol{X} := \boldsymbol{V}(M:=K)$ and $\boldsymbol{Y} := \boldsymbol{W}(N:=L)$ for the image-to-text (I2T)  version, and $\boldsymbol{X} := \boldsymbol{W}(M:=L)$ and $\boldsymbol{Y} := \boldsymbol{V}(N:=K)$ for the text-to-image (T2I) version. ``:='' is an assignment operator. $\boldsymbol{V}\in \mathbb{R}^{K*D}$, $\boldsymbol{W}\in \mathbb{R}^{L*D}$ are defined in Section 2.1, and will be assigned to $\boldsymbol{X}$ or $\boldsymbol{Y}$ in different retrieval versions.

Instead of treating all region-word pairs equally, a multi-head cross-modal attention mechanism was adopted to allocate weights to regions and words based on their contribution.
\begin{align}
\boldsymbol{Y}^{*} &=MultiHead(\boldsymbol{Query}:=\boldsymbol{X},\boldsymbol{Key}:=\boldsymbol{Y},\boldsymbol{Value}:=\boldsymbol{Y}) \nonumber \\
&=Concat(\boldsymbol{head}_1,...,\boldsymbol{head}_i,...,\boldsymbol{head}_H)\boldsymbol{Z}^O
\end{align}

where $\boldsymbol{Y}^{*} \in \mathbb{R}^{M*D}$,  $\boldsymbol{Z}^O \in \mathbb{R}^{D*D}$, $H$ is the number of attention heads, and $Concat({\cdot})$ represents the concatenation operation across the feature dimension $D$. In this work, $\boldsymbol{head}_i = Att(\boldsymbol{Query}_i:=\boldsymbol{X}{\boldsymbol{Z}_i}^{X}, \boldsymbol{Key_i}:= \boldsymbol{Y}{\boldsymbol{Z}_i}^{Y},\boldsymbol{Value_i}:=\boldsymbol{Y}{\boldsymbol{Z}_i}^{Y})$, where $Att$ refers to the scaled dot-product attention, and $\boldsymbol{Z}_i^{X}$,$\boldsymbol{Z}_i^{Y} \in \mathbb{R}^{D*\frac{D}{H}}$.

For I2T, the output $\boldsymbol{Y}^{*}$ denotes the region-attended text representations $\boldsymbol{W}^* = [\boldsymbol{w}^*_1,\boldsymbol{w}^*_2,...,\boldsymbol{w}^*_K] \in \mathbb{R}^{K*D}$ for each region, while for T2I, it refers to the word-attended image representations $\boldsymbol{V}^* =[\boldsymbol{v}^*_1,\boldsymbol{v}^*_2,...,\boldsymbol{v}^*_L] \in \mathbb{R}^{L*D}$ for each word. Here, $\boldsymbol{w}^*_i \in \mathbb{R}^{D}$ represents the region-attended text representation for the $i$-th region, and $\boldsymbol{v}^*_i \in \mathbb{R}^{D}$ represents the word-attended image representation for the $i$-th word. 

\textbf{Global Image-Text Fusion.} 
To further capture complex cross-modal correlation, we jointly mapped holistic images and sentences into a common space for global semantic consistence and heterogeneity minimization. To be specific, for the region-attended text representations $\boldsymbol{W}^*$ obtained in the process of local fusion, we first computed its average vector(denoted as $\overline{\boldsymbol{W}}^* = \frac{\sum_{i=0}^{K}\boldsymbol{w}_i^*}{K}$), and then projected it into a common space. The output result is $\boldsymbol{W_g}=\boldsymbol{X_w}^T \overline{\boldsymbol{W}}^*$, which serves as the final global representation of text, and $\boldsymbol{X_w}$ is a learnable embedding matrix. Similarly, we separately projected the average vector of $\boldsymbol{V}^*$ (represented as $\overline{\boldsymbol{V}}^*= \frac{\sum_{i=0}^{L}\boldsymbol{v}_i^*}{L}$) and the feature vector of the entire image (represented as $\boldsymbol{G}$) into the common embedding space as $\boldsymbol{V_{g_1}} = \boldsymbol{X_v}^T \boldsymbol{\overline{V}^*}$ and $\boldsymbol{V_{g_2}} = \boldsymbol{X_g}^T \boldsymbol{G}$. $\boldsymbol{X_v}$ and $\boldsymbol{X_g}$ are all learnable embedding matrixes. The final global representation of image can be fused as $\boldsymbol{V_{g}} = Fusion(\boldsymbol{V_{g_{1}}},\boldsymbol{V_{g_{2}}})$. 

\subsection{AsCL for I-T Matching}
The matching score between image $I$ and text $T$ consists of two components: local similarity score and global similarity score. According to the region-attended text representations $\boldsymbol{W}^*$ and the word-attended image representations $\boldsymbol{V}^*$, the local matching score is formulated as, 
\begin{equation}
S_{local}(I,T) = \frac{\sum_{i=1}^{k} R( \boldsymbol{v}_{i}, \boldsymbol{w}_{i}^{*})}{2K} + \frac{\sum_{j=1}^{L}R( \boldsymbol{w}_{j}, \boldsymbol{v}_{j}^{*})}{2L}
\end{equation}
, where $R( \boldsymbol{x}, \boldsymbol{y} ) = \frac{ \boldsymbol{x}^{T} \boldsymbol{y} }{||\boldsymbol{x}|| \cdot ||\boldsymbol{y}||}$. Meanwhile, the global matching score between image $I$ and text $T$ is represented as:
\begin{equation}
 S_{global}(I,T) = R(\boldsymbol{V_g},\boldsymbol{W_g}).
\end{equation}

Based on the local and global matching scores, the similarity score between image $I$ and text $T$ can be defined as:
\begin{equation}
    S(I,T)=u_1 \cdot S_{local}(I,T) + (1-u_1) \cdot S_{global}(I,T)
    \label{Eq:similarity}
\end{equation}
where $u_1$ is a hyper-parameter.

On this basis, we proposed a novel asymmetry-sensitive contrastive learning method by exploiting the generated positives and negatives according to our defined asymmetry types.

\textbf{I-T Matching for Asymmetry-\uppercase\expandafter{\romannumeral1}.} During training, there are $N$ image-text pairs in a batch. According to our defined Asymmetry-\uppercase\expandafter{\romannumeral1}, we generated $N$ negative sentences by adding noise. For each positive pair $(I, T)$, we retrieved $N-1$ in-batch negative images $\{\widehat{I_n}\}_{n=1}^{N-1}$, $N-1$ in-batch negative sentences $\{\widehat{T_n}\}_{n=1}^{N-1}$, and especially $N$ generated negative sentences $\{T^{-}_n\}_{n=1}^{N}$ for Asymmetry-\uppercase\expandafter{\romannumeral1}. Therefore, the objective function for Asymmetry-\uppercase\expandafter{\romannumeral1} can be formulated as follows: 
\begin{align}\label{Eq:NCEN}
    &L_{\uppercase\expandafter{\romannumeral1}}(I,T) =
    \frac{e^{S(I,T)/\tau}}{e^{S(I,T)/\tau}+\sum_{n=1}^{N-1}e^{S(\widehat{I_{n}},T)/\tau}} \\
    &+ \frac{e^{S(I,T)/\tau}}{e^{S(I,T)/\tau}+\sum_{n=1}^{N-1}e^{S(I,\widehat{T_{n}})/\tau}+  \sum_{n=1}^{N} \alpha_{n} \cdot e^{S(I,T^{-}_{n})/\tau}} \nonumber
\end{align}
where $\alpha_{n}$ is set to zero if the similarity score of the image-text pair exceeds the positive pair, otherwise we set $\alpha_{n}$ to 1.  

\textbf{I-T Matching for Asymmetry-\uppercase\expandafter{\romannumeral2}\&\uppercase\expandafter{\romannumeral3}.} 
For each positive image-text pair $(I, T)$, we generated the content-variant long sentence or short sentence as another positive sample, $T^+$, according to our defined Asymmetry-\uppercase\expandafter{\romannumeral2} and Asymmetry-\uppercase\expandafter{\romannumeral3}. Here, $(I,T^+)$ is a newly generated positive pair, while $T^+$ is also considered as a negative sample for other images in the same batch. Likewise, we added noise information to textual representations of these generated positives as additional negative samples. Consequently, there exist $N-1$ in-batch negative images $\{\widehat{I_n}\}_{n=1}^{N-1}$, $N-1$ in-batch negative sentences $\{\widehat{T^+_n}\}_{n=1}^{N-1}$, and $N$ generated negative sentences $\{\widetilde{T^-_n}\}_{n=1}^{N}$ towards every positive pair $(I,T^+)$. For Asymmetry-\uppercase\expandafter{\romannumeral2} \& \uppercase\expandafter{\romannumeral3}, $T^+$ plays a similar role in Equation \ref{Eq:NCEP} as $T$ in Equation \ref{Eq:NCEN}.

\begin{align}\label{Eq:NCEP}
&L_{\uppercase\expandafter{\romannumeral2}\&\uppercase\expandafter{\romannumeral3}}(I,T^+) =
    \frac{e^{S(I,T^+)/\tau}}{e^{S(I,T^+)/\tau}+\sum_{n=1}^{N-1}e^{S(\widehat{I_{n}},T^+)/\tau}} \\
    &+\frac{e^{S(I,T^+)/\tau}}{e^{S(I,T^+)/\tau}+\sum_{n=1}^{N-1}e^{S(I,\widehat{T^+_{n}})/\tau}+\sum_{n=1}^{N} \alpha_{n} \cdot e^{S(I,\widetilde{T^-_{n}})/\tau}} \nonumber
\end{align}

\textbf{Overall Training.} The overall training of AsCL for all types of information asymmetry is shown as follows:
\begin{equation}
    L_{AsCL}(I,T) = \frac{1}{2} L_{\uppercase\expandafter{\romannumeral1}}(I,T) + \frac{1}{2}L_{\uppercase\expandafter{\romannumeral2}\&\uppercase\expandafter{\romannumeral3}}(I,T^+)
    \label{Eq:Loss}
\end{equation}

\section{Experiments and Analyses}
\subsection{Experimental Settings}
\textbf{Dataset.} We conducted experiments on two benchmark datasets: MSCOCO \cite{MSCOCO} and Flickr30K \cite{F30K}. (1) MSCOCO: It consists of  a train split of 113,287 images, a validation split of 5,000 images, and a test split of 5,000 images. Each image is annotated with 5 sentences. We adopted the evaluation setting MSCOCO (5K), \textit{i.e.}, directly testing on the full 5K images. (2) Flickr30K: It contains 31,783 images, each with 5 corresponding sentences. It was split into 29,783 training images, 1,000 validation images and 1,000 testing images.

\textbf{Implementation Details.} We adopted R@K(K=1,5,10) to evaluate performance, which measures the percentage of ground truth being retrieved among top-K results. Higher R@K indicates better performance. We optimized AsCL on one NVIDIA Tesla A100 using PyTorch library. The Adam optimizer was employed with a batch size 64 and 20 epochs. We set the dimension of joint embedding space $D$ to 1024, the number of regions $K$ to 36. For MSCOCO, the learning rate was set to 5e-4 with decaying 10\% of every 10 epochs, $\tau$ was set to 0.05 and $u_1$ was set to 0.8. For Flickr30K, the learning rate was set to 2e-4 at first and declined by ten times every 10 epochs, $\tau$ was set to 0.01 and $u_1$ was set to 0.6. 

\subsection{Image-Text Retrieval Results}
As shown in Table \ref{Performance-Comparison}, our method outperformed existing state-of-art baselines across two benchmark datasets for the image-text retrieval task. For MSCOCO (5K), compared with LexLIP \cite{LexLip}, our model achieved a huge improvement of 24.6\% (94.8\% vs. 70.2\%) in terms of I2T/R@1 and 13.5\% (66.7\% vs. 53.2\%) in terms of T2I/R@1. For Flickr30K (1K), our model outperformed LexLIP by 7.7\% (99.1\% vs. 91.4\%) on I2T/R@1 and 4.6\% (83.0\% vs. 78.4\%) on T2I/R@1. Overall, according to fine-grained information asymmetry types, AsCL exhibits great effectiveness and superiority for the image-text retrieval task by leveraging corresponding generated samples. 

\begin{table*}[ht]
\tiny
\centering
\caption{Comparison results of the image-text retrieval on MSCOCO (5K test set) and Flickr30K (1K test set). We conducted experiments 5 times and reported the average value.}
\label{Performance-Comparison}
\resizebox{\textwidth}{!}{
\begin{tabular}{|l | c c c | c c c | c c c | c c c|}
\hline
\multirow{3}{*}{\textbf{Model}} & \multicolumn{6}{c|}{\textbf{MSCOCO (5K test set)}} & \multicolumn{6}{c|}{\textbf{Flickr30K (1K test set)}}\\
\cline{2-13} 

& \multicolumn{3}{c|}{I2T} & \multicolumn{3}{c|}{T2I}  & \multicolumn{3}{c|}{I2T} & \multicolumn{3}{c|}{T2I} \\
\cline{2-13}  

& R@1 & R@5 & R@10 & R@1 & R@5 & R@10 & R@1 & R@5 & R@10 & R@1 & R@5 & R@10  \\
\hline

VSE++ \cite{VSE} & 41.3 & 71.1 & 81.2 & 30.3 & 59.4 & 72.4 & 52.9 & 80.5 & 87.2 & 39.6 & 70.1 & 79.5 \\

SCAN \cite{SCAN} & 50.4 & 82.2 & 90.0 & 38.6 & 69.3 & 80.4 & 67.4 & 90.3 & 95.8 & 48.6 & 77.7 & 85.2  \\

IMRAM \cite{IMRAM} & 53.7 & 83.2 & 91.0 & 39.6 & 69.1 & 79.8 & 74.1 & 93.0 & 96.6 & 53.9 & 79.4 & 87.2 \\

DIME \cite{DIME} & 59.3 & 85.4 & 91.9 & 43.1 & 73.0 & 83.1 & 81.0 & 95.9 & 98.4 & 63.6 & 88.1 & 93.0 \\

DSRAN \cite{DSRAN} & 57.9 & 85.3 & 92.0 & 41.7 & 72.7 & 82.8 & 80.5 & 95.5 & 97.9 & 59.2 & 86.0 & 91.9\\

TAGS-DC \cite{TAGS} & 67.8 & 89.6 & 94.2 & 53.3 & 80.0 & 88.0 & 90.6 & 98.8 & 99.1 & 77.3 & 94.3 & 97.3 \\  

UNITER+DG \cite{Loss8} & 51.4 & 78.7 & 87.0 & 39.1 & 68.0 & 78.3 &  78.2 & 93.0 & 95.9 & 66.4 & 88.2 & 92.2 \\

SOHO \cite{SOHO} & 66.4 & 88.2 & 93.8 & 50.6 & 78.0 & 86.7 & 86.5 & 98.1 & 99.3 & 72.5 & 92.7 & 96.1\\

ViSTA \cite{Vista} & 68.9 & 90.1 & 95.4 & 52.6 & 79.6 & 87.6 & 89.5 & 98.4 & 99.6 & 75.8 & 94.2 & 96.9 \\

COTS \cite{COTS} & 69.0 & 90.4 & 94.9 & 52.4 & 79.0 & 86.9 &  90.6 & 98.7 & 99.7 & 76.5 & 93.9 & 96.6  \\

LightningDoT \cite{LightningDOT} & 64.6 & 87.6 & 93.5 & 50.3 & 78.7 & 87.5 & 86.5 & 97.5 & 98.9 & 72.6 & 93.1 & 96.1 \\

LexLIP \cite{LexLip} & 70.2 & 90.7 & 95.2 & 53.2 & 79.1 & 86.7 & 91.4 & 99.2 & 99.7 & 78.4 & 94.6 & 97.1 \\
\hline

\textbf{AsCL (OURS)} & \textbf{94.8} & \textbf{99.4} & \textbf{99.8} & \textbf{66.7} & \textbf{87.3} & \textbf{92.3} & \textbf{99.1} & \textbf{99.8} & \textbf{100} & \textbf{83.0} & \textbf{95.4} & \textbf{97.6}\\ 

\hline
\end{tabular}}
\end{table*}

\subsection{Ablation Experiments and Results}
\textbf{Different Samples.} We compared three variant models with different generated samples according to different asymmetry types: (1) ``\textit{-w/o} Pos.'': We only generated negative sentences for Asymmetry-\uppercase\expandafter{\romannumeral1}. (2) ``\textit{-w/o} Neg.'': We only generated positive sentences for Asymmetry-\uppercase\expandafter{\romannumeral2} and Asymmetry-\uppercase\expandafter{\romannumeral3}. (3) ``\textit{-w/o} P\&N'': We did not generate additional samples. As shown in Table \ref{Pos-Neg}, after removing generated positives, generated negatives and all generated samples, the results on Flickr30K (1K) decreased gradually, from 99.1\% to 98.8\% on I2T/R@1, from 99.8\% to 99.4\% on I2T/R@5, from 83.0\% to 76.4\% on T2I/R@1 and from 95.4\% to 91.5\% on T2I/R@5. Similar results can be observed on the MSCOCO (5K). The performance degradation verified the necessity of generated samples instantiated from different asymmetry types. 

\begin{table}[ht]
\centering
\caption{Ablation Results on MSCOCO and Flick30K.}
\label{Pos-Neg}
\resizebox{\linewidth}{!}{ 
\begin{tabular}{| l | c c | c c | c c | c c|}
%\toprule
\hline
\multirow{3}{*}{\textbf{Model Variants}}  & \multicolumn{4}{c|}{\textbf{MSCOCO (5K)}} & \multicolumn{4}{c|}{\textbf{Flickr30K (1K)}} \\
\cline{2-9} 

   & \multicolumn{2}{|c}{I2T} & \multicolumn{2}{|c|}{T2I} & \multicolumn{2}{c}{I2T} & \multicolumn{2}{|c|}{T2I} \\
\cline{2-9} 

 & R@1 & R@5 & R@1 & R@5 & R@1 & R@5 & R@1 & R@5 \\
\hline
\textbf{AsCL}  & 94.8 & 99.4  & 66.7 & 87.3  & 99.1 & 99.8  & 83.0 & 95.4 \\

\quad \textit{-w/o} Pos. & 93.1 & 99.1  & 65.1 & 86.0 & 99.0 & 99.7  & 79.6 & 92.7 \\

\quad \textit{-w/o} Neg. & 94.8 & 99.3  & 61.2 & 83.3  & 98.8 & 99.6  & 77.2 & 91.9 \\

\quad \textit{-w/o} P\&N & 93.1 & 98.8  & 59.1 & 81.6  & 98.8 & 99.4  & 76.4 & 91.5 \\

%\quad \textit{-w} only EDA  & 81.8 & 95.9 & 39.3 & 62.0 & 98.6 & 99.0  & 52.4 & 72.0   \\

\quad \textit{-w/o} MF   & 39.7 & 71.2 & 31.0 & 58.7 & 77.5 & 95.9 & 47.0 & 71.3 \\

\quad \textit{-w} triplet loss & 89.4 & 99.8 & 58.6 & 79.8 & 98.3 & 99.2 & 73.3 & 87.9    \\
\hline
\end{tabular}
}
\end{table}

\begin{figure}[ht]
	\centering
	\begin{minipage}{1\linewidth}
		\subfigure[Positives: Asymmetry-\uppercase\expandafter{\romannumeral2}\&\uppercase\expandafter{\romannumeral3}]{
			\label{fig:Diverse-Positives}
			\includegraphics[width=0.49\linewidth]{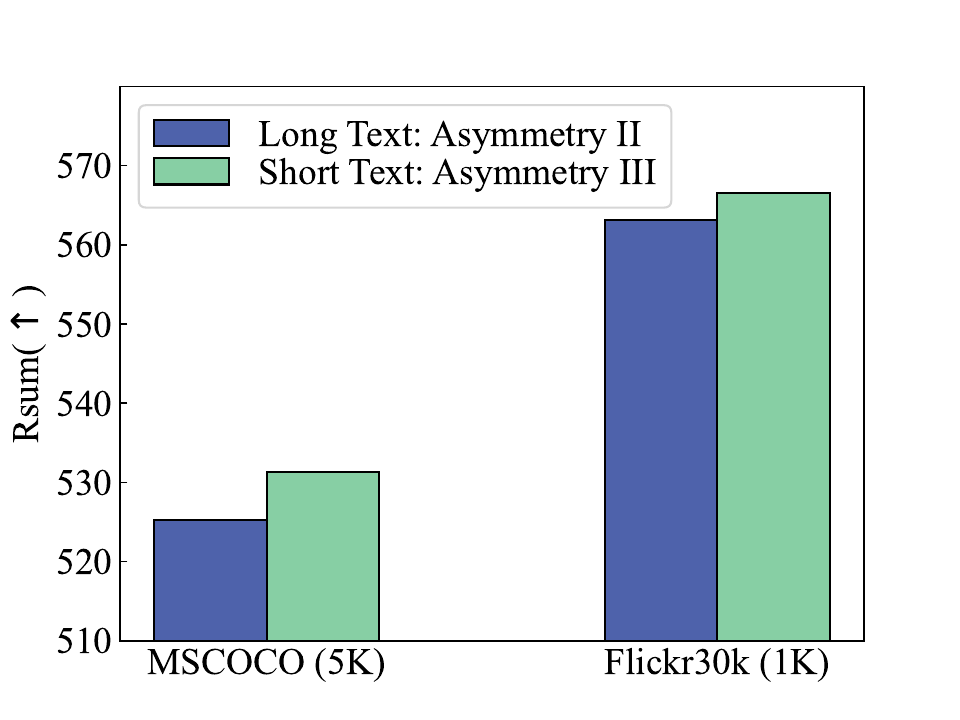}	
		}\noindent
  %Diverse Negatives: Asymmetry-\uppercase\expandafter{\romannumeral1}
		\subfigure[Negatives: Asymmetry-\uppercase\expandafter{\romannumeral1}]{
			\label{fig:Diverse-Negatives}
			\includegraphics[width=0.49\linewidth]{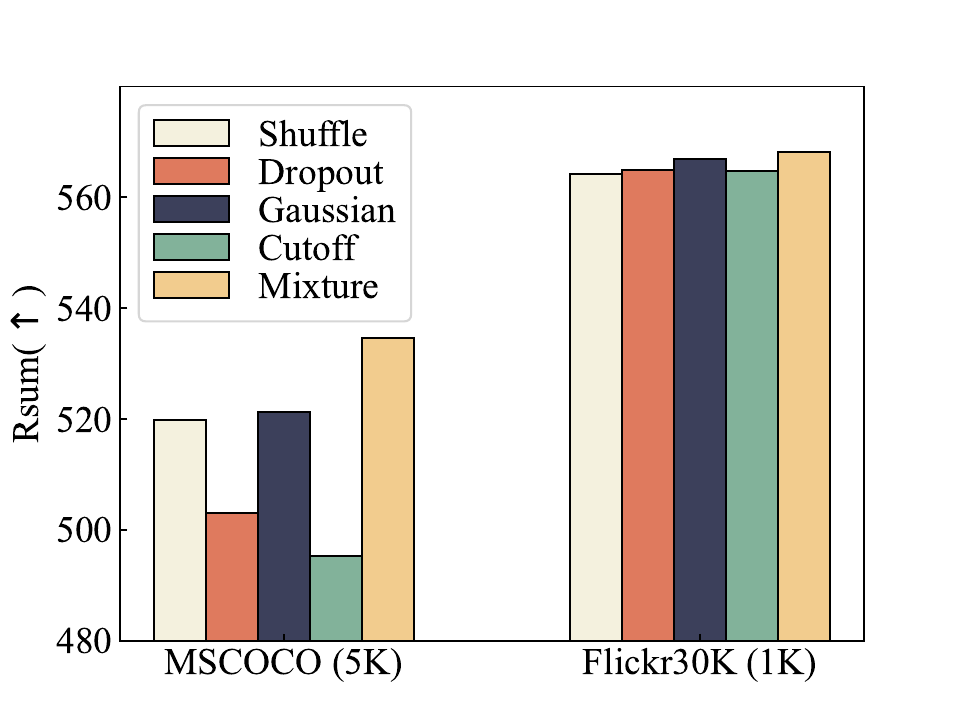}
		}
	\end{minipage}
	\caption{Performance comparison of diverse samples based on fine-grained generation strategies for each asymmetry type. (a) Influence of different generated positives according to Asymmetry-\uppercase\expandafter{\romannumeral2} and Asymmetry-\uppercase\expandafter{\romannumeral3}. (b) Influence of generated negatives with different noise addition strategies according to Asymmetry-\uppercase\expandafter{\romannumeral1}. $\uparrow$ means higher, better.}
	\label{fig:Diversity}
\end{figure}

We delved into more fine-grained sample generation strategies for each asymmetry type. The generation of negatives employs five noise addition strategies for Asymmetry-\uppercase\expandafter{\romannumeral1}, including ``Shuffle'', ``Dropout'', ``Gaussian'', ``Cutoff'' and ``Mixture''. The generation of positives involves two strategies, concatenation for Asymmetry-\uppercase\expandafter{\romannumeral2} and truncation for Asymmetry-\uppercase\expandafter{\romannumeral3}. As illustrated in Figure \ref{fig:Diversity}, we found that: (1) The short text for Asymmetry-\uppercase\expandafter{\romannumeral3} outperformed the generated long text for Asymmetry-\uppercase\expandafter{\romannumeral2}. We argued that although overlapping concepts between the short text and the given image has been reduced, the short text for Asymmetry-\uppercase\expandafter{\romannumeral3} should still be considered as a positive sample, which helps provide more diversity perspectives for learning robust semantic representation. (2) The Rsum values of ``Shuffle'', ``Dropout'', ``Gaussian'', and ``Cutoff'' were all lower than that of ``Mixture''. We supposed that ``Mixture'' enriches the diversity of negative textual descriptions for Asymmetry-\uppercase\expandafter{\romannumeral1}, thereby improving retrieval performance. Rsum is the sum of R@1+R@5+R@10 in both I2T and T2I. Higher Rsum means better performance.

\textbf{Cross-modal Fusion.} After removing cross-modal fusion (denoted as ``\textit{-w/o} MF'' in Table \ref{Pos-Neg}), all metrics on two datasets dramatically decreased, indicating that the hierarchical modal fusion via multi-head attention mechanism helps discover latent alignment from both global and local perspectives.

\textbf{Objective Function.} Compared to AsCL, the performance of triplet loss (denoted as ``\textit{-w} triplet loss'' in Table \ref{Pos-Neg}) declined, with R@1/I2T from 94.8 to 89.4 and R@1/I2T from 66.7 to 58.6 on MSCOCO (5K), which verified the superiority of our proposed asymmetry-sensitive contrastive learning. 

\subsection{Alignment and Uniformity Experiments and Results}
Alignment and uniformity are two key properties related to the quality of contrastive learning, which play an important role in the task of image-text retrieval.

\begin{figure}[ht]
	\centering
	\begin{minipage}{1\linewidth}
		\subfigure[Alignment: Positives get close.]{
			\label{fig:Positives-Distance}
			\includegraphics[width=0.49\linewidth]{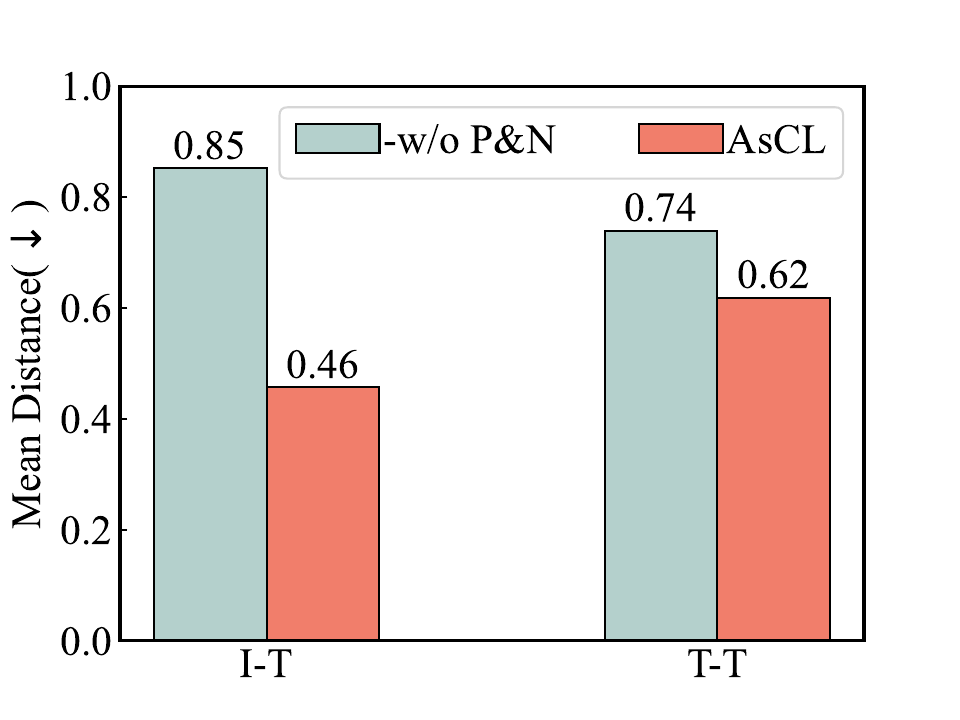}	
		}\noindent
		\subfigure[Uniformity: Negatives get apart.]{
			\label{fig:Negatives-Distance}
			\includegraphics[width=0.49\linewidth]{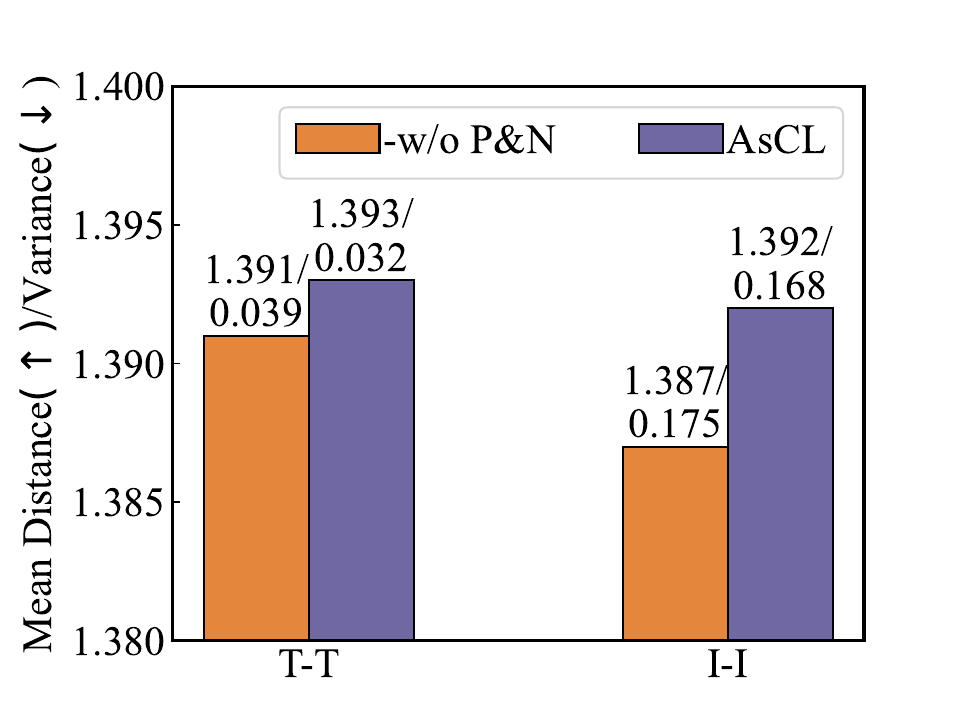}
		}
	\end{minipage}
	\caption{Mean distance between positive pairs(a) and negative pairs(b) in high-dimensional space on MSCOCO. $\uparrow$ means higher, better. $\downarrow$ means lower, better.}
	\label{fig:Distance}
\end{figure}

\textbf{Alignment Evaluation.} Alignment prefers a closer distance between positive pairs. Positive pairs consists of two parts: inter-modal image-text pairs and intra-modal text-text pairs. We separately calculated the average Euclidean distance between positive pairs in MSCOCO (5K) under our model AsCL and the ablated model AsCL$_{-w/o \thinspace P\&N}$. As shown in Figure \ref{fig:Positives-Distance}, compared with representations learned from model AsCL$_{-w/o \thinspace P\&N}$, both the average distance between positive image-text (I-T) pairs and the average distance between positive text-text (T-T) pairs learned from AsCL decreased. These results indicated that our method acquired high-quality semantic representations, thereby achieving better inter-modality and intra-modality semantic alignment.

\textbf{Uniformity Evaluation.} Uniformity favors a uniform distribution of feature embeddings on a hypersphere. We calculated the average Euclidean distance among all images and all texts with different semantics in MSCOCO (5K), respectively. In Figure \ref{fig:Negatives-Distance}, compared with model AsCL$_{-w/o \thinspace P\&N}$, the average spacing among different texts under AsCL enlarged (1.393 vs 1.391) , along with the drop of variance (0.032 vs 0.039). The average distance and variance between different images exhibited similar phenomenon. It indicated that representations learned from AsCL are more widely and evenly distributed, which verified better uniformity.

\subsection{Impact on Text Queries with Different Lengths}
\begin{figure}[ht]
	\centering
	\begin{minipage}{1\linewidth}
		\subfigure[Image Retrieval in MSCOCO.]{
			\label{fig:MSCOCO-Length}
			\includegraphics[width=0.49\linewidth]{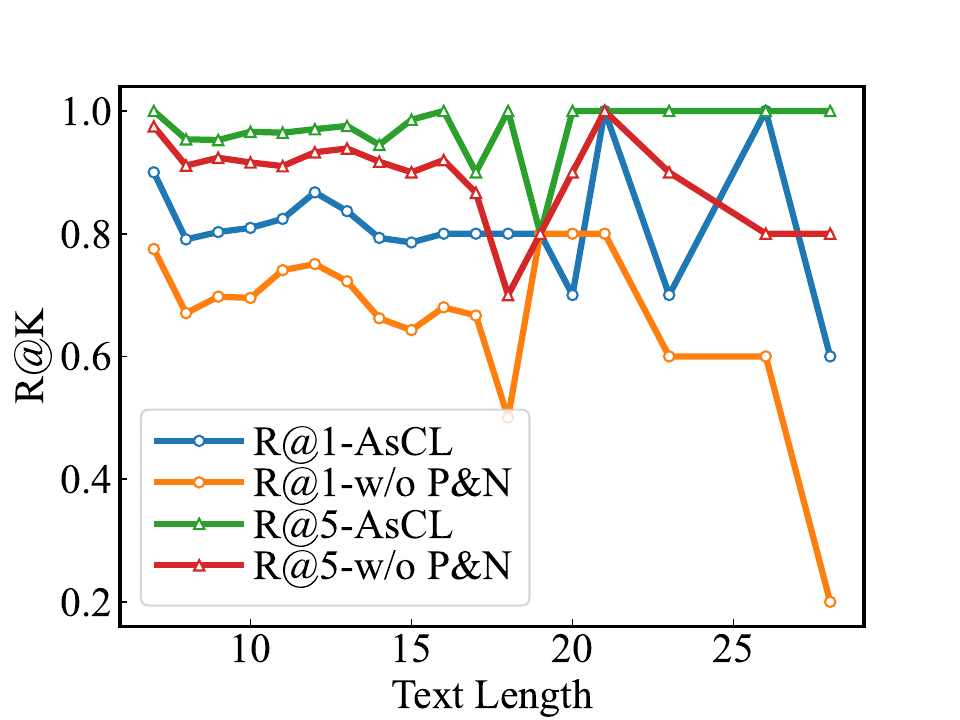}	
		}\noindent
		\subfigure[Image Retrieval in Flickr30K.]{
			\label{fig:Flickr30K-Length}
			\includegraphics[width=0.49\linewidth]{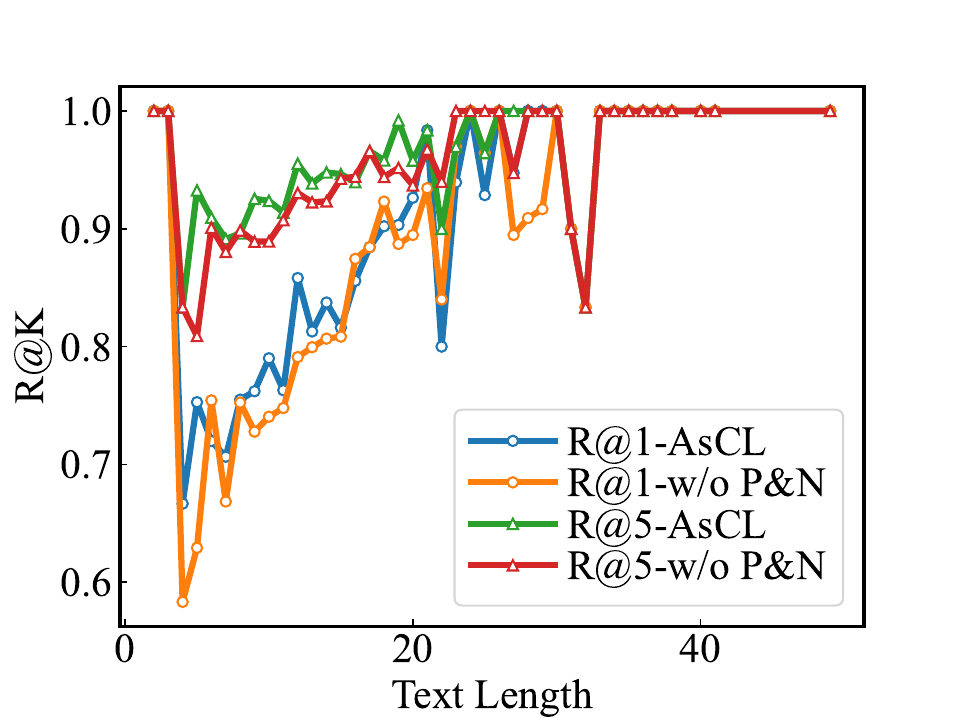}
		}
	\end{minipage}
	\caption{Image retrieval for text queries of different lengths.}
	\label{fig:Text-Length}
\end{figure}

We further conducted text-to-image(T2I) task based on textul queries with different text lengths. We sampled 1,000 sentences from MSCOCO and Flickr30K, respectively. From Figure \ref{fig:Text-Length}, for those textual queries (especially less than 10 words), AsCL yielded better image retrieval results. One possible explanation is that we generated short sentences as positive samples based on Asymmetry-\uppercase\expandafter{\romannumeral3} during training, which makes AsCL more sensitive and effective to short sentences.

\section{Related Work}
Previous work for image-text retrieval mainly focused on unified representation and cross-modal interaction. These methods can be categorized into two types: (1) dual encoder \cite{VSE,DSRAN,LBPS,TFS}, where images and text are separately encoded into a common embedding space by two individual encoders; (2) cross encoder \cite{Oscar,ViLBERT,UNITER}, where images and texts are jointly encoded by a cross encoder architecture, incorporating two heterogeneous modalities into unified forms. 

For the image-text retrieval task, most recent approaches \cite{SCAN,IMRAM,COTS} employ contrastive loss with hard negatives, \textit{i.e.}, samples that are more difficult to be distinguished. The optimization process of contrastive loss with hard negatives increases the similarity score of positive image-text pairs while decreasing that of negative pairs, thereby achieving better retrieval performance. Hence, selection strategies for informative samples have been extensively explored. Early works \cite{Loss6,Loss7} randomly chose negatives from the original dataset for training. Subsequently, based on the generation methods, researchers have incorporated more difficult negatives into contrastive learning. In the UNITER+DG \cite{Loss8} framework, hard negative sentences were sampled based on structure relevance by using a denotation graph. Fan et al. \cite{TAGS} proposed TAGS-DC to generate synthetic sentences automatically as negative samples. Filip Radenovic et al. \cite{Loss10} presented an importance-sampling approach that reweighted negative samples within a batch, aiming to upsample harder negatives in proportion to their difficulty. Based on fine-grained information asymmetry types, our paper proposed an asymmetry-sensitive contrastive learning method with generated positives and negatives for the image-text retrieval task.

\section{Conclusion}
In this paper, we presented a novel asymmetry-sensitive contrastive learning method. Concretely, in order to address fine-grained information asymmetry issues, we generated corresponding positives and negatives for each asymmetry type, which are fully utilized in the optimization of contrastive learning. Our approach enhances sensitivity to subtle differences between two heterogeneous modalities and achieves more discriminative multimodal semantic representations for the subsequent image-text retrieval task. Moreover, from both local and global perspectives, a hierarchical cross-modal fusion module was proposed to capture sophisticated correspondence between visual and semantic modalities through the multimodal attention mechanism. Experimental results on two widely used datasets, \textit{i.e.}, MSCOCO and Flickr30K, have verified that our method outperforms previous state-of-the-art baselines.
% In this paper, in order to address fine-grained information asymmetry issues, we presented a novel asymmetry-sensitive contrastive learning method with hierarchical cross-modal fusion, which helps enhance sensitivity to subtle differences between two heterogeneous modalities and achieves more discriminative multimodal semantic representations for the subsequent image-text retrieval task. Experimental results on two widely used datasets have verified that our method outperforms previous state-of-the-art baselines.

\section*{Acknowledgment}
This work is support by the National Natural Science Foundation of China (NO. U1811461, 61572250), the Jiangsu Province Science \& Technology Research Grant (BE2021729), Key R\&D Program Project of Nanjng Jiangbei New Area (ZDYF20200130), and the Collaborative Innovation Center of Novel Software Technology and Industrialization, Jiangsu, China.

\bibliographystyle{IEEEbib}
\bibliography{icme2023template}
% \bibitem{b1} G. Eason, B. Noble, and I. N. Sneddon, ``On certain integrals of Lipschitz-Hankel type involving products of Bessel functions,'' Phil. Trans. Roy. Soc. London, vol. A247, pp. 529--551, April 1955.
% \bibitem{b2} J. Clerk Maxwell, A Treatise on Electricity and Magnetism, 3rd ed., vol. 2. Oxford: Clarendon, 1892, pp.68--73.
% \bibitem{b3} I. S. Jacobs and C. P. Bean, ``Fine particles, thin films and exchange anisotropy,'' in Magnetism, vol. III, G. T. Rado and H. Suhl, Eds. New York: Academic, 1963, pp. 271--350.
% \bibitem{b4} K. Elissa, ``Title of paper if known,'' unpublished.
% \bibitem{b5} R. Nicole, ``Title of paper with only first word capitalized,'' J. Name Stand. Abbrev., in press.
% \bibitem{b6} Y. Yorozu, M. Hirano, K. Oka, and Y. Tagawa, ``Electron spectroscopy studies on magneto-optical media and plastic substrate interface,'' IEEE Transl. J. Magn. Japan, vol. 2, pp. 740--741, August 1987 [Digests 9th Annual Conf. Magnetics Japan, p. 301, 1982].
% \bibitem{b7} M. Young, The Technical Writer's Handbook. Mill Valley, CA: University Science, 1989.
% \end{thebibliography}
\vspace{12pt}
\color{red}
% IEEE conference templates contain guidance text for composing and formatting conference papers. Please ensure that all template text is removed from your conference paper prior to submission to the conference. Failure to remove the template text from your paper may result in your paper not being published.

\end{document}